\documentclass[useAMS,usenatbib]{mn2e}

\usepackage{amsmath,amssymb,graphicx}

\newcommand{\kepler}{\textit{Kepler}}

\title[TTV vs. RV]{Sensitivity bias in the mass-radius distribution from Transit Timing Variations and Radial Velocity measurements}
\author[Steffen et al.]{
Jason H. Steffen$^{1}$
\\
$^{1}$University of Nevada, Las Vegas, Department of Physics and Astronomy,\\
4505 S. Maryland Parkway, Box 454002, Las Vegas, NV 89154-4002
}

\begin{document}


\pagerange{\pageref{firstpage}--\pageref{lastpage}} 

\maketitle

\label{firstpage}

\begin{abstract}
Motivated by recent discussions, both in private and in the literature, we use a Monte Carlo simulation of planetary systems to investigate sources of bias in determining the mass-radius distribution of exoplanets for the two primary techniques used to measure planetary masses---Radial Velocities (RVs) and Transit Timing Variations (TTVs).  We assert that mass measurements derived from these two methods are comparably reliable---as the physics underlying their respective signals is well understood.  Nevertheless, their sensitivity to planet mass varies with the properties of the planets themselves.  We find that for a given planet size, the RV method tends to find planets with higher mass while the sensitivity of TTVs is more uniform.  This ``sensitivity bias'' implies that a complete census of TTV systems is likely to yield a more robust estimate of the mass-radius distribution provided there are not important physical differences between planets near and far from mean-motion resonance.  We discuss differences in the sensitivity of the two methods with orbital period and system architecture, which may compound the discrepancies between them (e.g., short period planets detectable by RVs may be more dense due to atmospheric loss).  We advocate for continued mass measurements using both approaches as a means both to measure the masses of more planets and to identify potential differences in planet structure that may result from their dynamical and environmental histories.
\end{abstract}


\section{Introduction}

The detection of transiting exoplanets, coupled with dynamical measurements of their masses, allows us to determine the relationship between a planet's mass and its size---especially in regimes where there are no solar system analogs.  Two means to determine planet masses dynamically have been employed in recent discoveries.  Radial Velocity (RV) measurements have been the longstanding pillar of this effort and have produced most of the relevant mass determinations.  Transit Timing Variations (TTVs) have come to the scene more recently and are enabled by the nearly continuous observations of NASA's \kepler\ mission \citep{Borucki:2010}.  As new mass measurements continue to enlarge our sample of characterized planets, differences in the planet properties as determined by the two methods have become increasingly apparent.

\begin{figure}
\includegraphics[width=0.49\textwidth]{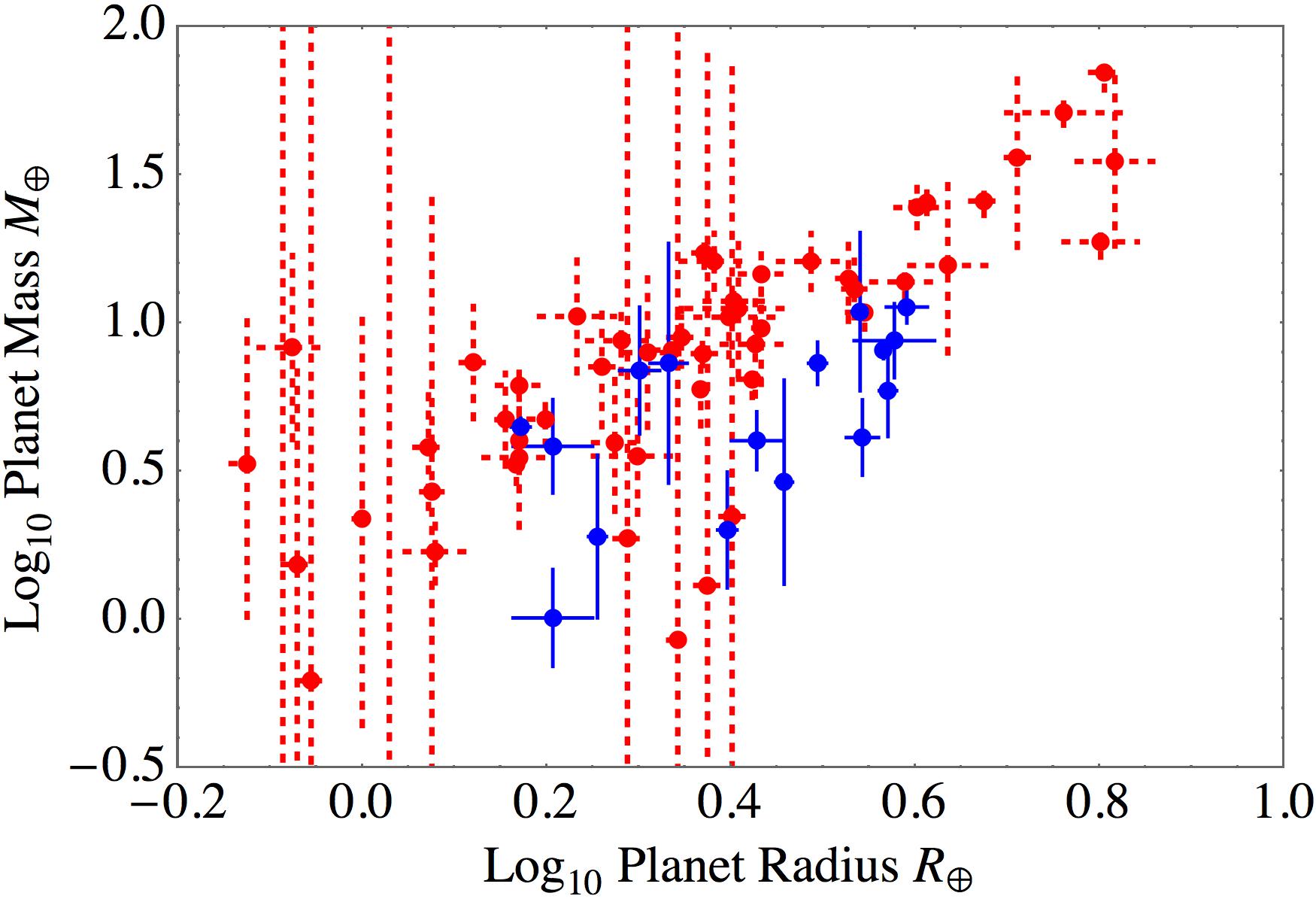}
\caption{Observed mass and radius data as reported in \citet{Wolfgang:2015} with TTV measured planets shown as solid blue and RV measured planets as dotted red.  Note that the TTV planets are systematically lower than the RV planets.\label{observedmr}}
\end{figure}

Figure \ref{observedmr} shows the values of the mass and radius for many relevant planets (the data are taken directly from \citet{Wolfgang:2015}).  The red (dashed) points in this figure are planets characterized by RV data while the blue (solid) points are determined from TTVs.  We see from this picture that the typical TTV planet has a systematically lower density than its RV counterparts (falling below and/or to the right of the RV planets).  Initially, there was concern that the measurements from one or the other method may be incorrect (TTVs being the newer approach, was the most suspicious) and discussions have arisen both in private conversations and in the literature regarding this apparent discrepancy \citep[e.g.,][]{Weiss:2014,Jontof-Hutter:2014,Dai:2015}.  That being said, however, analyses of both TTV and RV data on the Kepler-11 system and the WASP-47 system show that the two methods yield the same results \cite{weisstalk,Dai:2015}.

Besides, fundamentally both TTVs and RVs are manifestations of dynamical interactions via gravity (a physical theory that is well vetted) so there is little reason to believe that parameter estimates using either method are biased in general.  If they were, it would require systematic noise variations that follow either the orbital periods of the planets, or the TTV modulation timescales, often for multiple planets in the same system---which is somewhat unlikely.  The main source of bias for the two methods may come from unseen planets contaminating the signal.  Given the narrow-band nature of TTV perturbations (observed planets with a given period ratio produce a signal at a very specific frequency), unseen planets may affect RV detections more significantly.  Multibody resonances that include an unseen planet, while rare, would have a strong effect on both methods.

We propose that the differences between the densities of planets measured by these two methods is not due to either method giving incorrect results.  Rather, these differences largely stem from the fact that the sensitivity of the two methods depends upon the physical and orbital properties of the planets themselves.  This ``sensitivity bias'' means that the two methods are better suited to probing different regions on the mass-radius plane.  Here we conduct a Monte Carlo simulation to determine the effects of this sensitivity bias as it relates to understanding the mass-radius relationship.  In Section \ref{SNR} we derive the dependence of the detection signal-to-noise ratio (SNR) for the two methods.  These expressions give insights into the results of our Monte Carlo simulation of exoplanet systems that is outlined in Section \ref{MC}.  In Section \ref{results} we present the results of our simulation---noting the differences in the planet properties that are most readily detected by each method.  Finally, in Section \ref{discussion} we discuss the implications of our simulations as well as possible caveats that merit further observational and theoretical study.

\section{Signal to noise ratios}\label{SNR}

The physics of both methods, gravitational dynamics, is well understood.  However, how that physics is manifest in both terms of the signal and the noise is different for the TTV and RV methods.  Nevertheless, in terms of the signal-to-noise ratio, the problems of extracting planet parameters are essentially the same---fitting a sine curve to noisy data.  In order to gain insight into the differences in sensitivity between the two methods we examine the important parameter dependencies of the SNR for each.

The RV signal (here $\Sigma_\text{RV}$) is essentially proportional to the mass of the planet and its orbital velocity about the host star.  The noise in the RV measurements (here $\sigma_\text{RV}$) comes from a combination of stellar, instrumental, environmental, and statistical fluctuations.  We will ignore the effects of planet multiplicity---they are issues for another time.  Thus, the SNR for RV data scales as:
\begin{eqnarray}
\text{SNR}_\text{RV} & \equiv & \frac{\Sigma_\text{RV}}{\sigma_\text{RV}} \\
& = & \frac{1}{\sigma_\text{RV}} \frac{2\pi a M_\text{p}}{P}\\ \label{RVSNR}
 & \sim & \frac{M_\text{p}}{\sigma_\text{RV} P^{1/3}},
\end{eqnarray}
where $a$ is the semi-major axis of the planet orbit, $P$ is its orbital period, and $M_\text{p}$ is the planet mass (units aren't particularly relevant here).  We used Kepler's law to remove the dependence on $a$ and dropped constant prefactors in the last step as our goal is to primarily to see how the SNR depends upon the properties of the planet in question and not to make quantitative statements.

An SNR calculation for the TTV signal is less straightforward.  Here the TTV signal is given by $\Sigma_\text{TTV}$ and the uncertainty in the transit time is given by $\sigma_0$ (we will use $\sigma_\text{TTV}$ later):
\begin{eqnarray}
\text{SNR}_\text{TTV} &\equiv & \frac{\Sigma_\text{TTV}}{\sigma_0}.
\end{eqnarray}
One important difference with the TTV SNR is that the timing uncertainty $\sigma_0$ depends upon the properties (size and orbital period) of the planet that is perturbed by the planet whose mass one is measuring.  That is, if you want to measure the mass of one planet, you study the transit times of the other planets in the system.  In addition, the TTV signal $\Sigma_\text{TTV}$ is a strong function of the orbital period ratio.  In order to proceed, we make the approximation that the orbital period and size of the perturbing planet are comparable to that of the transiting planet---a reasonable approximation for our purposes as we, again, are looking for a functional form to give insight rather than one to make quantitative statements.

Thus, the general form of the TTV signal is
\begin{eqnarray}
\Sigma_\text{TTV} = f(\mathcal{P}) M_\text{p} P
\end{eqnarray}
where $\mathcal{P}$ is the ratio of orbital periods and the function $f$ encapsulates the dependence on the period ratio.  This function $f$ itself scales approximately as $f \sim 1/ \Delta$ where $\Delta$ is a measure of distance from an MMR.  Issues related to this dependence on period ratio are beside the point of this work and interested readers are directed to the body of literature dealing with the nature of the TTV signal for details (e.g., \citet{Agol:2005,Holman:2005,Nesvorny:2008,Lithwick:2012a,Hadden:2015,Deck:2015}).  Therefore, we will drop the function $f$, yielding the straightforward dependence
\begin{eqnarray}
\Sigma_\text{TTV} \sim M_\text{p} P.
\end{eqnarray}

To separate the effects of the planet from those of the instrument, star, etc., we look more carefully at the parameters that contribute to the timing uncertainty.  \citet{Carter:2008} and \citet{Price:2014} show that, under appropriate circumstances (e.g., knowledge of the out-of-transit stellar flux and small impact parameter), the uncertainty in transit times is given by the expression
\begin{eqnarray} \label{noiseTTV}
\sigma_0 & = & \frac{1}{Q}\sqrt{\frac{\tau T}{2}}
\end{eqnarray}
where $\tau$ is the ingress and egress duration, $T$ is the transit duration, and $Q$ is
\begin{eqnarray}
Q & = & \frac{\delta}{\sigma}\sqrt{\Gamma T}.
\end{eqnarray}
In this expression for $Q$, $\Gamma$ is the sampling cadence, $\sigma$ encapsulates the fundamental statistical and instrumental noise properties, and $\delta$ is the depth of the transit signal:
\begin{eqnarray}
\delta & = & f_0 \left(\frac{R_\text{p}}{R_\star}\right)^2,
\end{eqnarray}
where $f_0$ is the out-of-transit flux from the star.

Beginning with Equation (\ref{noiseTTV}) we have:
\begin{eqnarray}
\sigma_0 & = & \frac{1}{Q}\sqrt{\frac{\tau T}{2}}\\
& = & \frac{\sigma}{\delta}\sqrt{\frac{\tau T}{2\Gamma T}}\\
& \sim & \frac{\sigma_\text{TTV}}{R_\text{p}^2}\sqrt{\tau}\\ \label{timingsigma}
& \sim & \frac{\sigma_\text{TTV}P^{1/6}}{R_\text{p}^{3/2}},
\end{eqnarray}
where $\sigma_\text{TTV}$ includes all of the factors that deal with the cadence, the star, and other systematic effects (recall that $\sigma_\text{TTV}$ is not the uncertainty in the transit time).  The last step uses the fact that the duration of ingress and egress are given by the planet size and orbital velocity ($\tau \sim R_\text{p} a/P = R_\text{p} P^{-1/3}$).  Note that, within the context of this analysis (which comes from the Fisher matrix) the timing uncertainty is independent of the transit duration $T$.  Finally, we use this last expression for $\sigma_0$ to arrive at the TTV SNR:
\begin{eqnarray}\label{TTVSNR}
\text{SNR}_\text{TTV} & \sim & \frac{M_\text{p}R_\text{p}^{3/2}P^{5/6}}{\sigma_\text{TTV}}.
\end{eqnarray}
(As before, factors of order unity have been dropped.)

Comparing Equations (\ref{TTVSNR}) and (\ref{RVSNR}) we see that a large planet radius or a longer orbital period can enable measurements of planets with smaller masses.  The different dependencies on orbital period implies that, for a given mass, RV planets will have systematically shorter orbits.  The difference in SNR (i.e., the sensitivity bias) for the two methods is partly responsible for the different densities of planets measured by them---rather than incorrect or biased mass measurements themselves.

\section{Monte Carlo}\label{MC}

In order to understand the effects of this sensitivity bias between the two measurment techniques, we conducted a Monte Carlo simulation of a large population of \kepler -like, two-planet systems.  Since most TTV signals are dominated by only one or (at most) two planets, the two-planet model is justified.  Even if two perturbers contribute a comparable amount, the signal will be at most $\sim \sqrt{2}$ larger and therefore would not make an appreciable difference for our purposes.  All of our observational data related to \kepler\ planet candidates comes from the Quarter 17 (Q17) catalog at the NASA Exoplanet Archive.

To construct a system we fix the stellar mass to one solar mass.  We draw random values for the initial orbital period, planet size, eccentricity, and longitude of pericenter for the inner planet.  The outer planet is then constructed using estimated distributions for planet orbital period ratios and size ratios.  Planet masses are assigned randomly using the mass-radius relationship given in \citep{Wolfgang:2015}:
\begin{equation}
\frac{M}{M_\oplus} \sim \text{Normal}\left( \mu = 2.7\left(\frac{R}{R_\oplus} \right)^{1/3}, \sigma = 1.9 \right).
\end{equation}

Eccentricities for all planets are Rayleigh distributed with a Rayleigh parameter of 0.026---which makes the eccentricity distribution comparable to rough estimates of the distribution of mutual orbital inclinations ($0.026 = 1.5^\circ \pi /180$).  We note that our goal is to produce a sample that is approximately consistent with the observed \kepler\ planet candidates and not to make definitive statements about the specific choices for those distributions or their parameter values.

\begin{figure}
\includegraphics[width=0.49\textwidth]{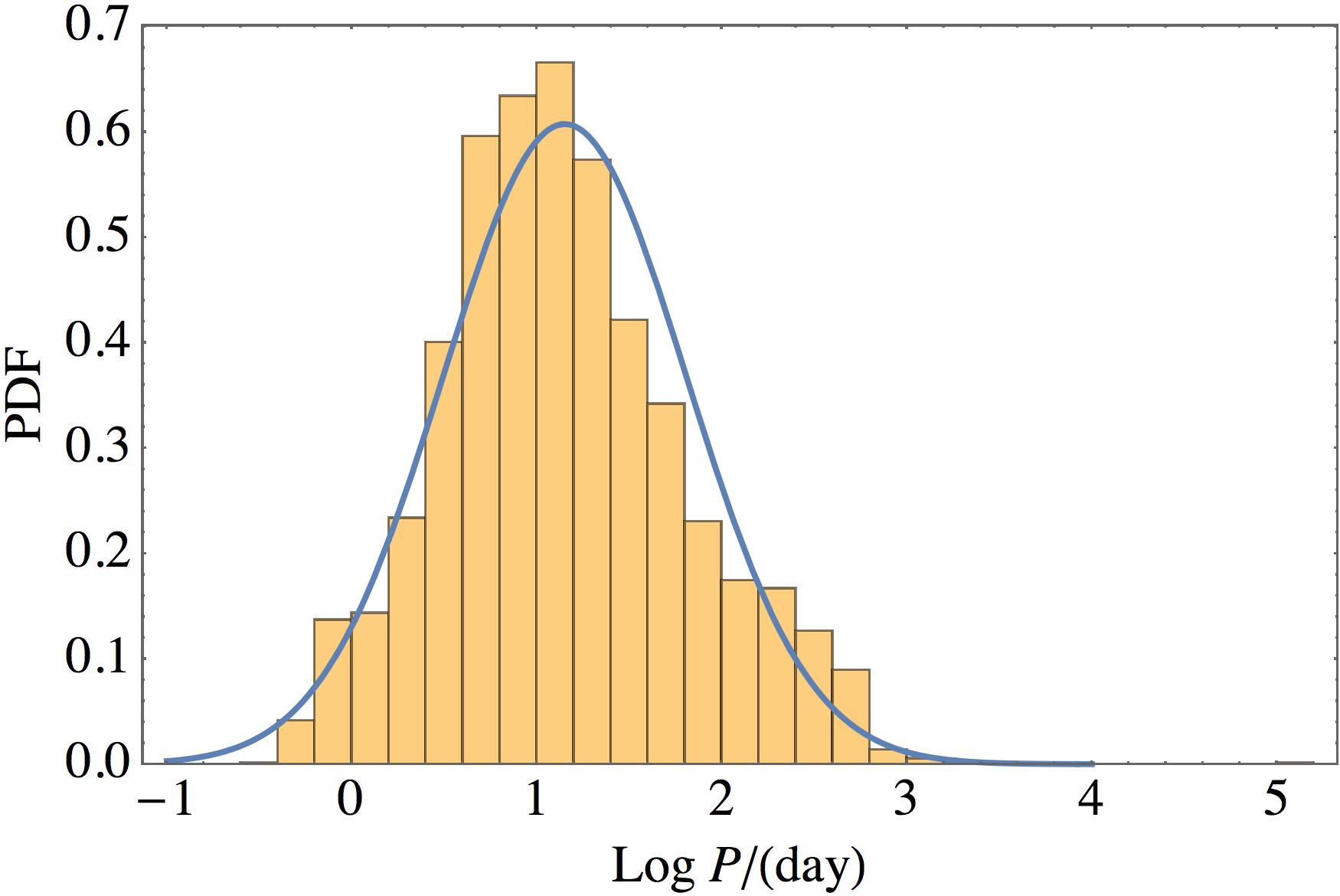}
\caption{Distribution of observed (histogram) and simulated (curve) orbital periods.  The model for the simulated distribution is log-normal with (log) mean 1.15 and standard deviation 0.654.  Realizations from this model with periods less than one day were discarded.\label{perdist}}
\end{figure}

The distribution of inner planet orbital periods comes from a maximum likelihood fit of the observed distribution of orbital periods for the \kepler\ planet candidates to a log-normal distribution.  There are a variety of reasons we choose not to use the distribution of only the innermost planets (e.g., we aren't interested in only the innermost planet pairs).  Regardless, the differences between the distribution of innermost planets and all planets is small---in part because the candidate list is dominated by single-planet systems.  Figure \ref{perdist} shows this distribution and our model, which has a log mean of 1.15 (14 days) and a log standard deviation of 0.654 (typical innermost planets have orbital periods between 3 and 63 days).  We drop all planets with orbital periods less than 1 day.

This last choice will have only a tiny effect on the differences between the RV and TTV methods in our simulation, but it does reflect a dichotomy between systems that are accessible to the different methods.  Specifically, the RV signal is larger at shorter orbital periods and there are examples of short-period RV planets from \kepler\ (e.g., Kepler-10 \citet{Batalha:2011}).  On the other hand, planets on such short orbital periods typically do not have nearby companions that would induce a large TTV signal \citet{Steffen:2013c}---to date there are no exceptions to this rule for solar-type stars through mid-M (Kepler-42 \citep{Muirhead:2012} is the lone counter example, but it is a very-late M near 0.1 $M_\odot$).

\begin{figure}
\includegraphics[width=0.49\textwidth]{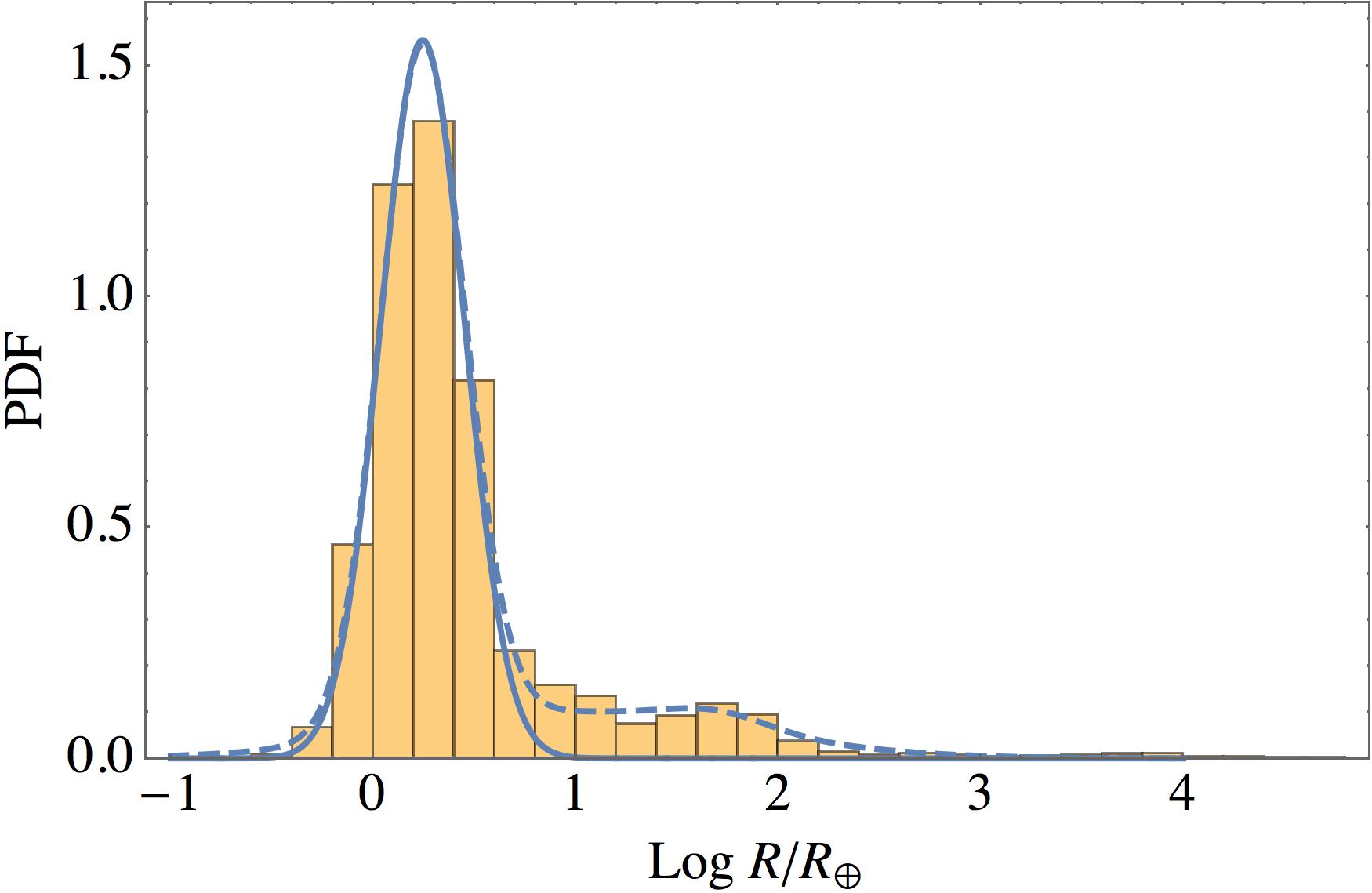}
\caption{Distribution of observed (histogram) and simulated (curves) planet sizes.  The model included three log normal distributions representing small planets, gas giants, and outlier signals.  The dashed curve represents the sum of the three distributions while the solid curve represents the contribution only from the small planets.  Monte Carlo realizations for the inner planet sizes were chosen from the small planet distribution which had a (log) mean of 0.244 and standard deviation of 0.208.\label{sizedist}}
\end{figure}

The distribution of inner planet sizes was similarly estimated from a maximum likelihood fit to the observed distribution of planet sizes.  Here the observed distribution was approximated by a mixture of three log-normal distributions---one that represents the typical, small \kepler\ planet, one that represents gas giants, and one that represents a sample of very large candidates (e.g., hundreds of Earth radii).  To facilitate the convergence of this model, only the two mixture ratios, the mean and standard deviation of the small planets,  and the standard deviation of the giant planets were allowed to float (the mean of the giant planet population was fixed at $10 R_\oplus$ and the log mean and log standard deviation of the outliers were fixed at 1.7 and 0.25 respectively).  The results of this fit are shown in Figure \ref{sizedist}.  For our simulation we use only the distribution that corresponds to the small planets with a log mean and standard deviation of  0.244 and 0.208 respectively.  We exclude from our simulation any planets smaller than $0.5 R_\oplus$.

\begin{figure}
\includegraphics[width=0.49\textwidth]{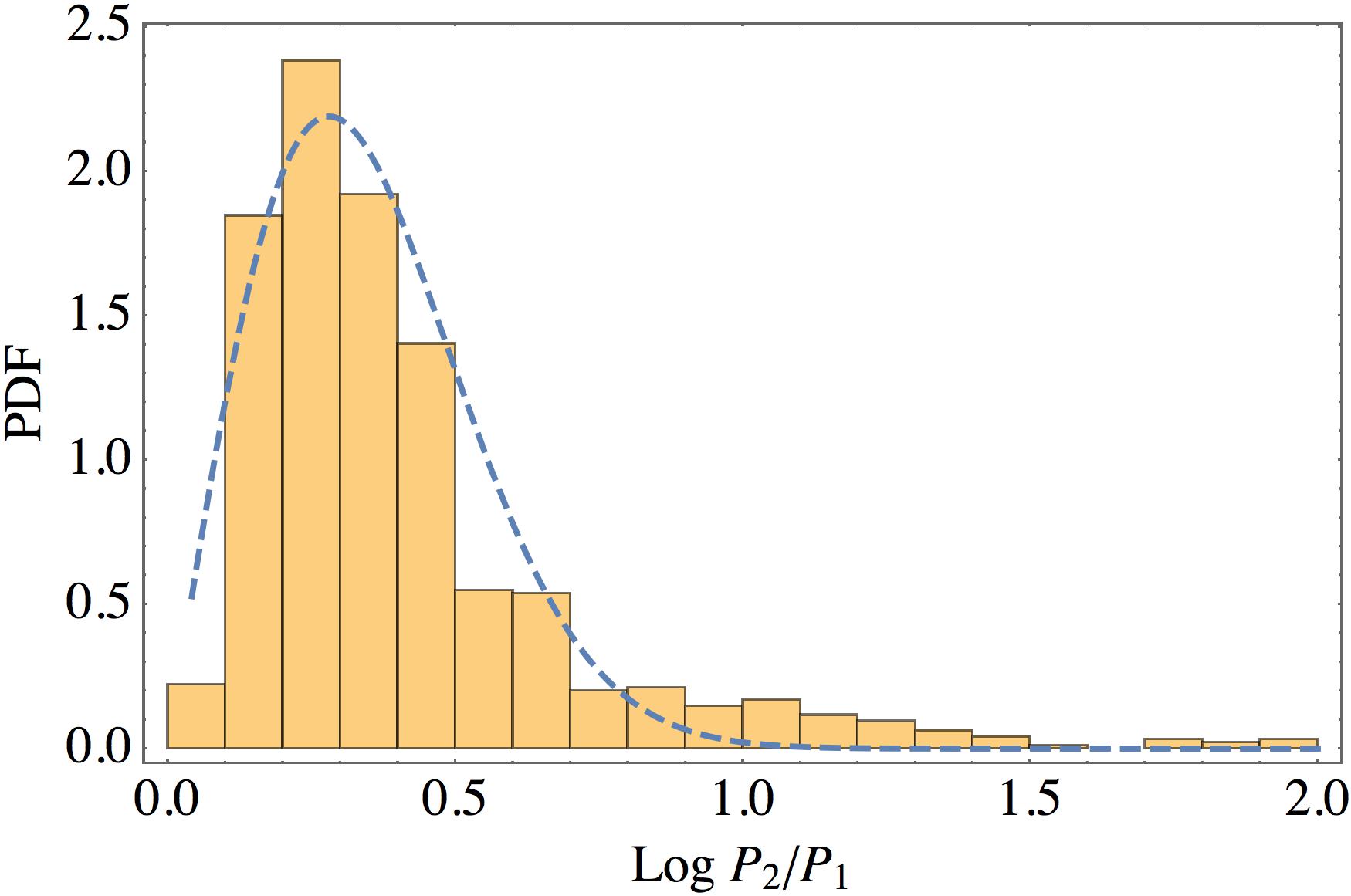}
\caption{Distribution of observed (histogram) and simulated orbital period ratios.  The distribution for simulated period ratios is taken from \citep{Steffen:2013c} and is not a fit to the data (indicated by a dashed rather than a solid curve)---unlike the other distributions.  This model is a Rayleigh distribution in log space, truncated to include values larger than 0.0414 (period ratios larger than 1.1) and with a Rayleigh parameter of 0.28.  The orbital period of the inner planet was chosen from the distribution in Figure \ref{perdist} and the orbital period of the outer planet was chosen using this distribution.  A consequence of this approach is that the distribution of outer planet orbital periods will be somewhat more broad than that of the inner planets (and hence more broad than the observed orbital periods).  This difference, however, will not make any substantive change to our conclusions.\label{pratdist}}
\end{figure}

The ratio of the orbital periods of the inner and outer planets were selected from the truncated Rayleigh distribution used as a model in \citep{Steffen:2013c}.  Specifically the period ratios must be larger than 1.1 (0.0414 in log space) and the (log) Rayleigh parameter is 0.28.  Figure \ref{pratdist} shows this distribution along with a histogram of the observed distribution of period ratios.  For this work, we did not fit a new period-ratio distribution (and for simplicity we did not use the bias-corrected distribution from \citep{Steffen:2015}).  We point to Figure \ref{pratdist} as evidence that the model we use is a reasonable approximation to what is observed.  Ultimately, the RV signal we consider will not be affected by this choice and the nature of the TTV signal will select planet pairs from small regions near Mean-Motion Resonance (MMR).  Thus, a substantial variety of smooth distributions would yield similar results.

\begin{figure}
\includegraphics[width=0.49\textwidth]{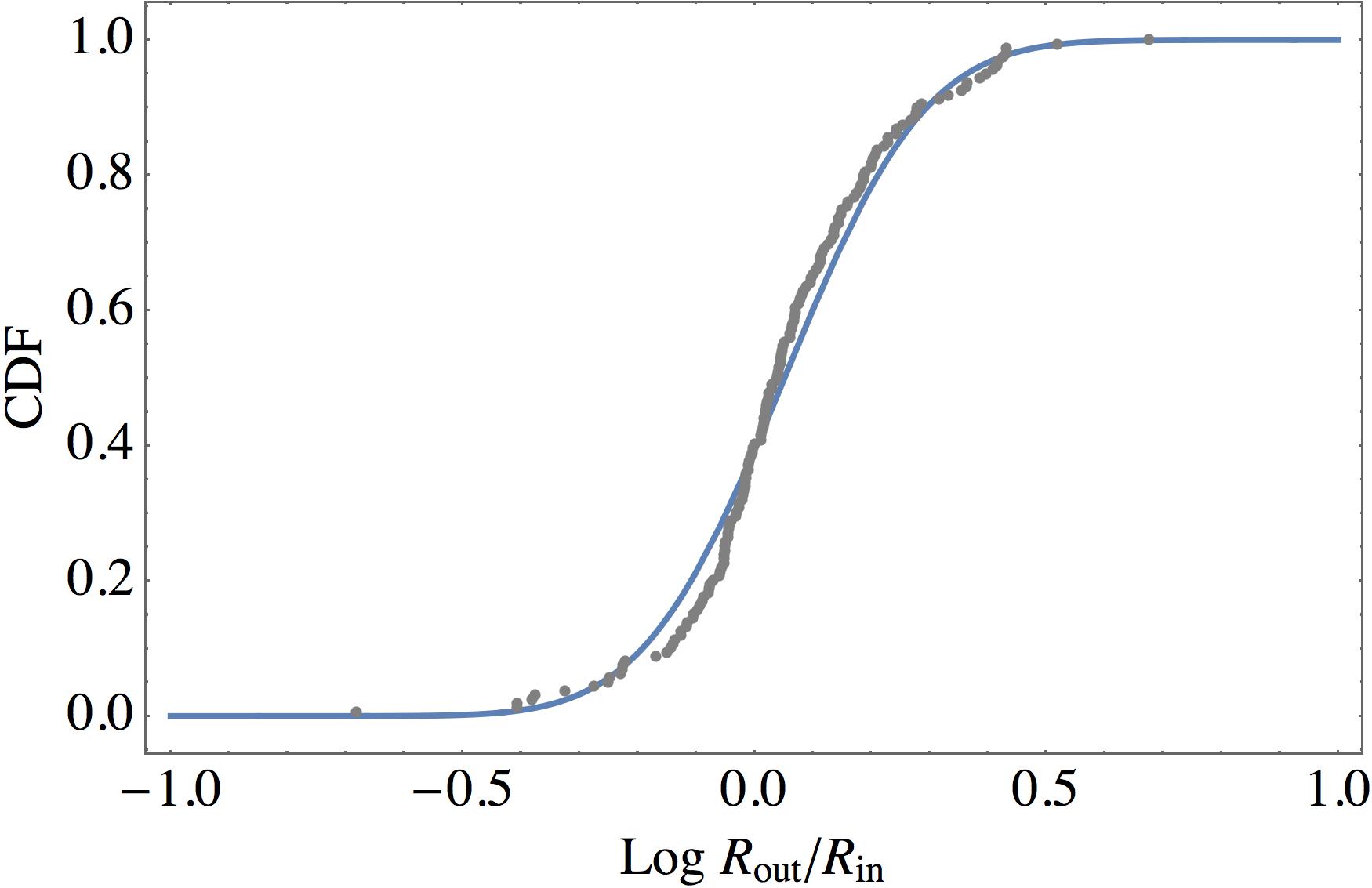}
\caption{Distribution of observed (data) and simulated planet size ratios.  The data were taken from \citep{Ciardi:2013} and the model is a fit to those data.  It is log normal with a (log) mean of 0.0514 and standard deviation of 0.190.  The size of the inner planet was drawn from the distribution in Figure \ref{sizedist} and the size of the outer planet was selected using this distribution.  A consequence of this approach is that the distribution of outer planet sizes will be slightly more broad than that of the inner planets (and hence more broad than the observed planets).  This difference, as with the period ratio distribution, will not have a substantive effect on our conclusions.\label{sizeratdist}}
\end{figure}

The size ratio of the inner planet to the outer planet was chosen using a fitted (log-normal) model to the distribution of size ratios given in \citet{Ciardi:2013}.  The fitted values for the log mean and standard deviations are 0.0514 and 0.190 respectively (i.e., the typical outer planet is 12\% larger than the inner planet, though the distribution is broad).  Figure \ref{sizeratdist} shows the data from \citet{Ciardi:2013} and our model fit.  Planet masses are assigned to each of the planets using the mass-radius distribution given in \citet{Wolfgang:2015}.

\begin{figure}
\includegraphics[width=0.49\textwidth]{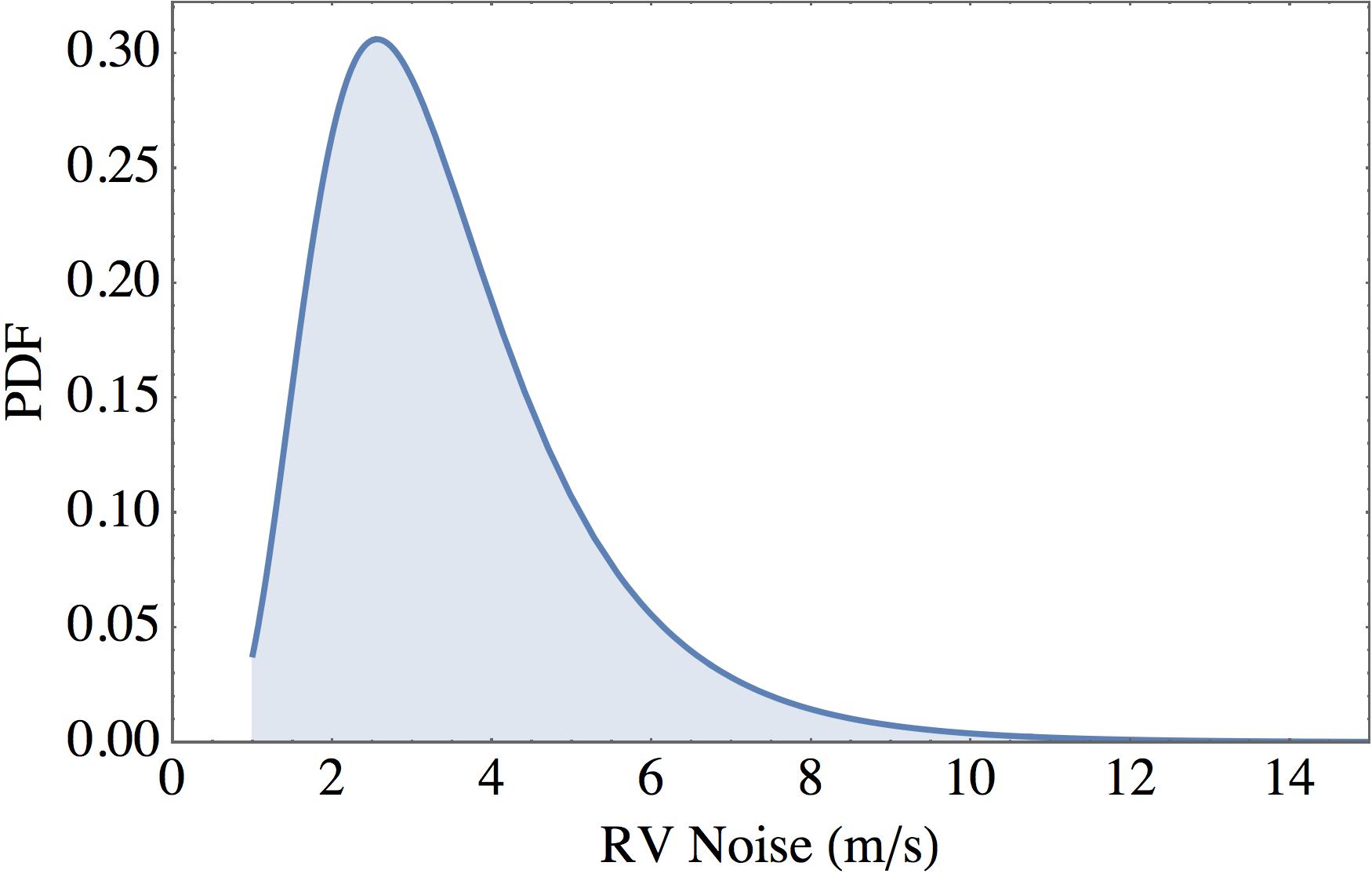}
\caption{RV precision distribution used for our Monte Carlo simulation.  This distribution is a truncated log normal distribution restricted to values larger than 0.5 m/s and with a (log) mean of 0.5 and standard deviation of 0.2.  The peak of the distribution is near 3.2 m/s.\label{rvsnr}}
\end{figure}

The RV precision that we assign to each planet is drawn from a log normal distribution truncated to be larger than 0.5 m/s.  The (log) mean and standard deviation of the distribution are 0.5 and 0.2 respectively---corresponding to a peak in the distribution near 3.2 m/s.  These parameters are not a fit to any existing data, but were chosen to approximately match the results seen in modern spectrographs when stellar noise and other systematic effects are included \citep{EPRVtalks}.  This distribution is shown in Figure \ref{rvsnr}.

\begin{figure}
\includegraphics[width=0.49\textwidth]{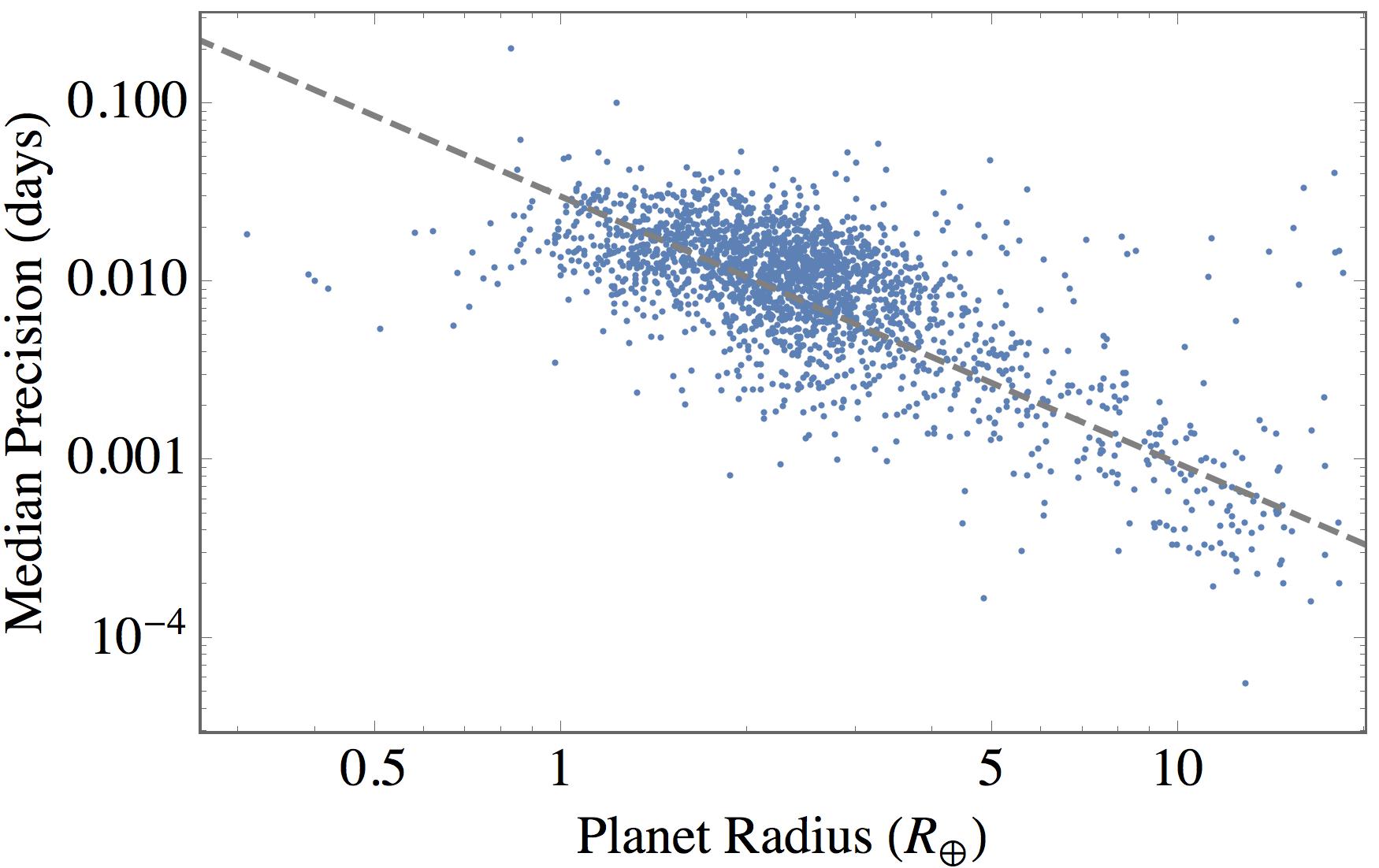}
\caption{Measured timing precision for \kepler\ planets as a function of planet radius as given in \citep{Holczer:2015}.  From these data we assigned 10 timing precision values to each of the planets in our TTV simulations.  Those 10 precisions were randomly selected from the 50 planets nearest in size to the simulated planet.  This approach captures many of the effects of stellar variability in determining the timing precision.  The dashed line (not a fit) shows the predicted dependence on timing uncertainty with $R^{-3/2}$ as shown in Equation (\ref{timingsigma}).\label{timingprecision}}
\end{figure}

Finally, there are a few other items we need in order to calculate the TTV signal for our sample systems.  First, from our initial sample of $10^4$ systems we randomly select 5000 for the TTV calculation.  We choose the time of first transit of the inner planet to be zero and the time of first transit of the outer planet to be a random number between zero and one orbital period of the outer planet.  We assume the system to be coplanar (mutual inclinations of $\sim 1.5^\circ$ would have a negligible effect on these results).  We also assign a timing precision for the transit midtimes.  This assignment is done using the median transit time uncertainties for the list of KOIs from \citet{Holczer:2015}.  We identify the 50 KOIs that are nearest in size to our simulated planet and randomly select the median timing uncertainty of 10 of them.  Thus, for each TTV system, we produce 10 SNR values.  Figure \ref{timingprecision} shows the median timing precision as a function of planet size from \citet{Holczer:2015}.

\section{Results}\label{results}

The TTV signal, which we define as the standard deviation of the timing residuals, is calculated for each planet using TTVFaster---analytic models given in \citep{Agol:2015}.  We identify as planet detections all individual planets where the TTV signal induced on the companion planet in the system is larger than two times its expected transit time uncertainty (recall that for each planet we compare the TTV signal to the measured timing uncertainty of 10 comparably-sized candidates).  In other words, if a given planet has a large TTV signal, we claim to have measured the mass of its perturbing planet.

\begin{figure}
\includegraphics[width=0.49\textwidth]{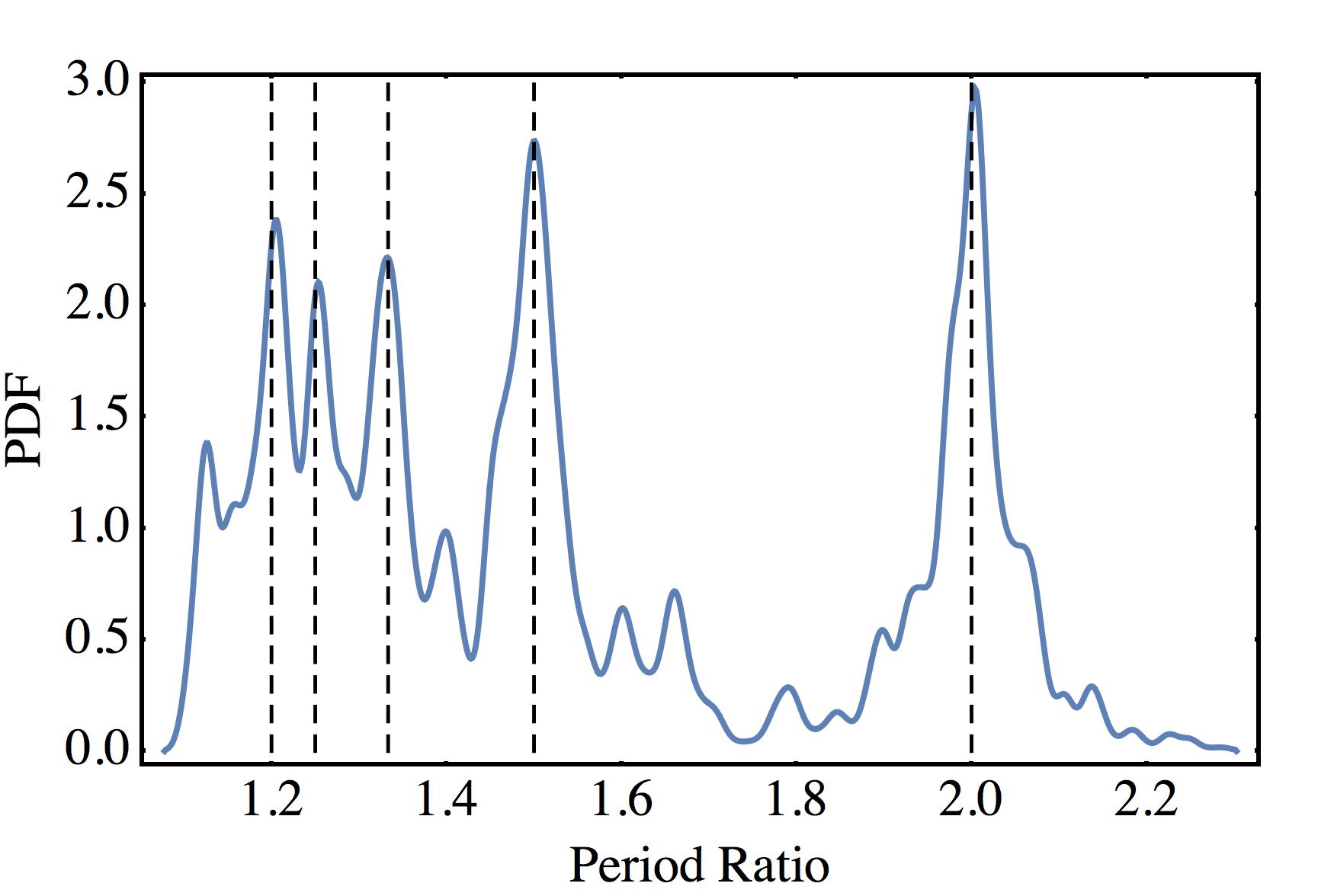}
\caption{A kernel density estimator distribution of period ratios with detected TTV planets (the bandwidth for this figure is 0.01).  There are peaks in this distribution near the expected, first-order MMR (dashed vertical lines).  Systems with TTV planets farther from resonance also tended to have slightly larger orbital periods---due to the fact that the TTV signal grows with orbital distance.\label{pratiosttv}}
\end{figure}

Since the TTV signal has a strong dependence on the ratio of orbital periods, we would expect that most TTV discoveries have period ratios near MMR.  The distribution of period ratios that resulted in TTV detections is shown in Figure \ref{pratiosttv}.  Indeed we see the peaks near first-order MMR.  Many of the planets that are somewhat far from MMR are either relatively close to higher order resonances, such as 5:3, or have longer orbital periods---which also increases the TTV SNR as shown in Equation (\ref{TTVSNR}).

Since our initially assigned eccentricities for the planets is quite small ($\sim 0.026$) we use the circular approximation for the RV signal for each planet in our sample.  We select as planet detections all of the individual planets with RV amplitudes larger than the assigned RV precision (i.e., an SNR of 1).  Note that we are choosing an SNR of 1 for RVs (where the RV ``$K$'' amplitude is the signal) and an effective SNR of 2 for the TTVs (where the standard deviation of the residuals is the signal).  Our conclusions are not dependent upon these, or other reasonable choices for what constitutes a ``detection''.  If we were to use an SNR of $\sqrt{2}$ for TTV detections our conclusions would generally be made stronger.

\begin{figure}
\includegraphics[width=0.49\textwidth]{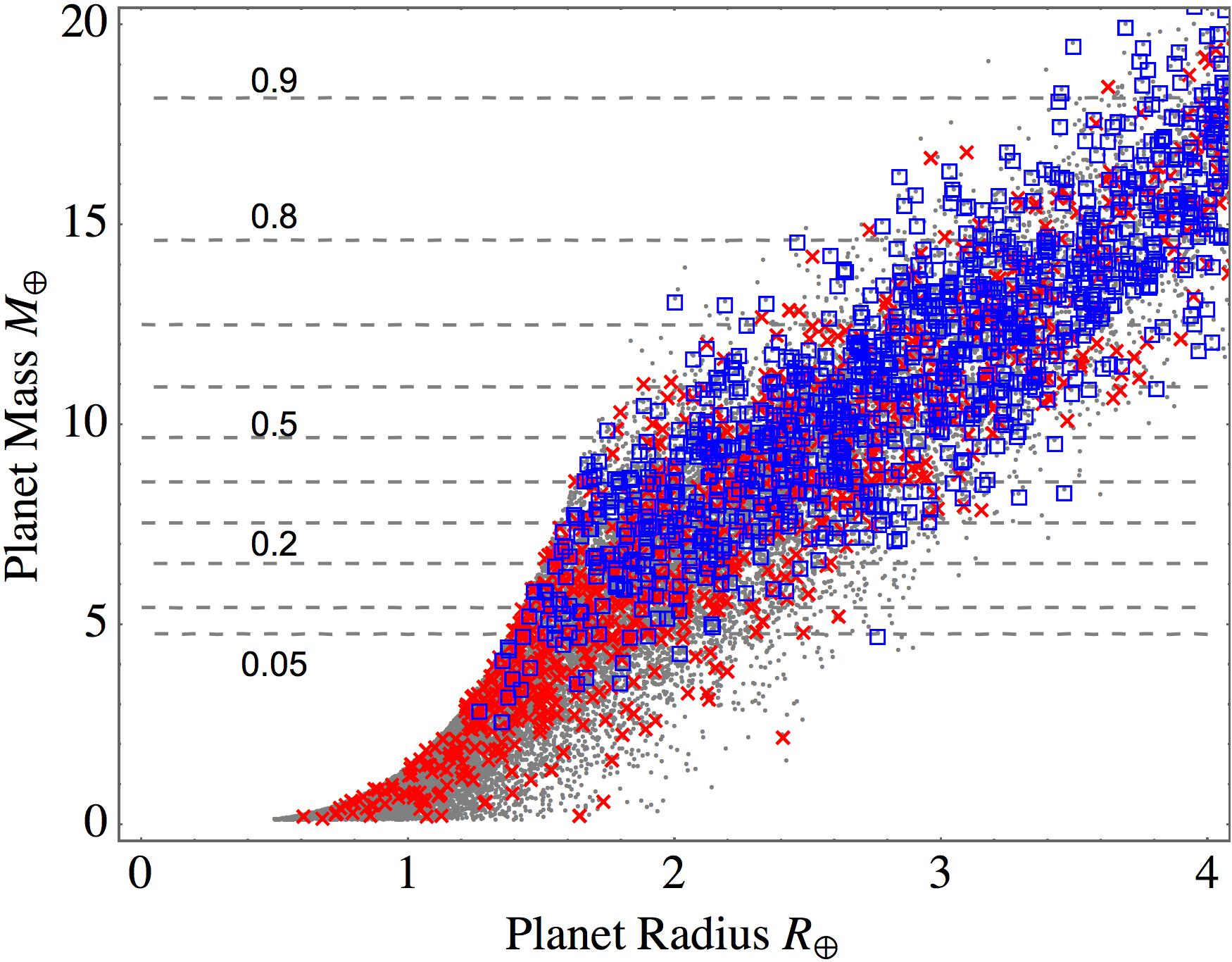}
\caption{Mass and radius of the simulated (gray dots) and a random sample of the detected planets using RV (blue squares) and TTVs (red crosses).  Only a subsample of the detected planets were selected for readability purposes.  Completeness contours derived from our input distributions (assuming a fixed 2.5 m/s RV noise---see text) are also shown.  One can see that below a certain size (here approximately 2.5 $R_\oplus$) the RV sample is biased toward more massive planets while TTVs can be sensitive to less massive planets of the same size.  We note that some regions of the plane (e.g., planets with particularly low or high density) are not probed by the mass-radius distribution from \citep{Wolfgang:2015}.\label{inputoutput}}
\end{figure}

Figure \ref{inputoutput} shows the planet mass as a function of radius for planets less than $4 M_\oplus$ for the initial simulated systems (gray dots) along with a random subsample of the RV and TTV detections (blue squares and red crosses respectively).  We show only a subsample for legibility purposes.  We see from this figure the underlying mass-radius distribution of \citet{Wolfgang:2015}---the clipped portion near one Earth radius is the Iron density cutoff ($M \sim R^{3.7}$).  We see that for a given planet size, TTVs are able to detect planets with lower masses.  This fact is responsible, at least partially, for the observation that planets characterized by TTVs have had systematically lower density than those measured by RVs.


Also shown in Figure \ref{inputoutput} are completeness contours calculated using our initial distributions---to simplify the calculation of these contours we used a fixed RV prevision of 2.5 m/s.  For a fixed RV prevision of this value, the deviations from the underlying mass-radius relationship for the detected planets ($M - \langle M \rangle$) is almost indistinguishable from the results of our full simulation where the RV precision is a random variable.  We see that there is a large rise in completeness throughout this range of planet masses.  For planets of Neptune mass and larger (Neptune size and larger), the completeness is quite good ($\gtrsim 80$\%), but for planets less than about 5 $M_\oplus$ (2 $R_\oplus$) the completeness is more than an order-of-magnitude smaller.

\begin{figure}
\includegraphics[width=0.49\textwidth]{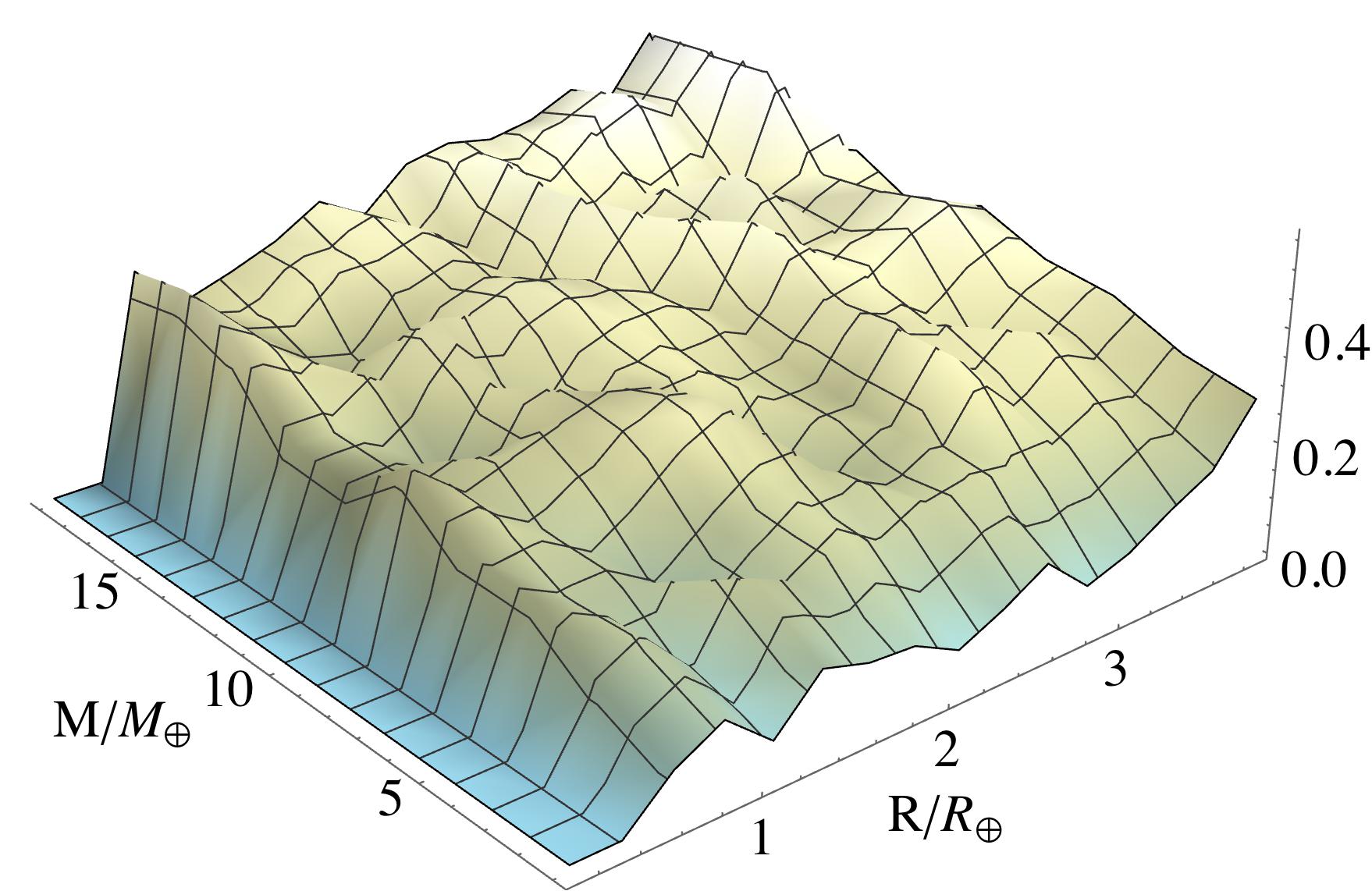}
\caption{The fraction of successful TTV detections as a function of planet mass and radius determined by Monte Carlo simulation.  We calculated the TTV signal for 20,000 planet pairs where the planet mass and size were uniformly distributed across this region of parameter space.  Aside from the increasing trend toward higher completeness as the planet size and mass grow, the completeness surface is relatively flat (variations much less than an order of magnitude, unlike RV sensitivity shown in Figure \ref{inputoutput}).  Due to the strong dependence of the TTV signal on period ratio, we caution against using this completeness calculation for precise quantitative estimates.\label{ttvcomplete}}
\end{figure}

A useful contrast is to see the completeness for TTV mass measurements in this same region of parameter space.  Figure \ref{ttvcomplete} shows the results of a completeness estimate using a Monte Carlo simulation of 20,000 planet pairs where planet sizes were selected randomly across the region from 0 to 20 $M_\oplus$ and from 0.5 to 4 $R_\oplus$.  Because TTV detections depend so strongly on the period ratio of the planets, these completeness estimates should be considered only qualitatively (especially since we did not use an initial period ratio distribution designed to reproduce the observed features near MMR).  Rather, the more important lesson from Figure \ref{ttvcomplete} is that the (somewhat noisy) completeness surface is relatively flat with a slight rise toward larger sizes and larger masses as expected---it does not change by orders-of-magnitude in the same way that RV completeness does.

\begin{figure}
\includegraphics[width=0.49\textwidth]{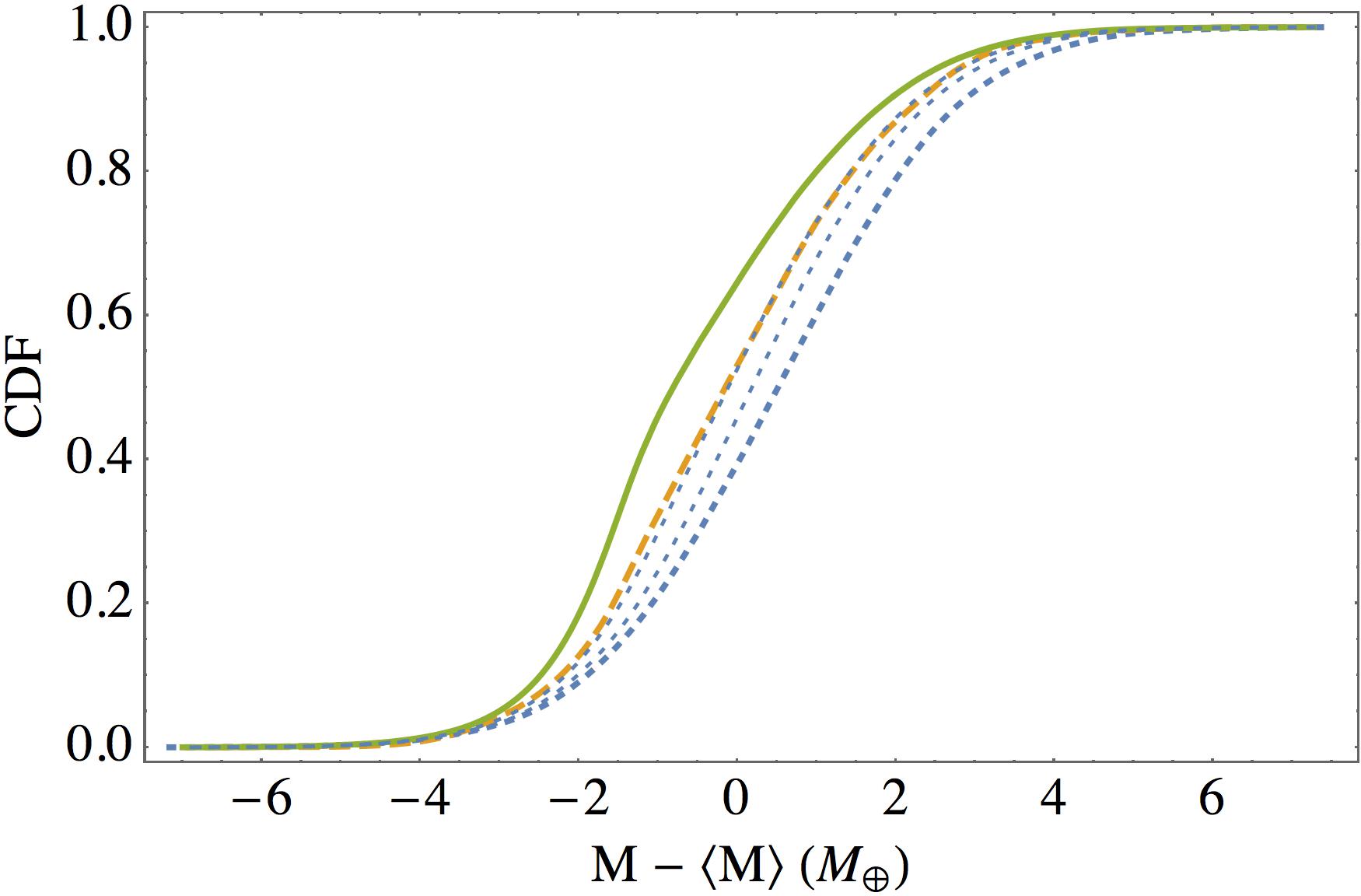}
\caption{Cumulative distribution of the simulated (green solid) and detected planets from our simulations (RV planets are blue dotted and TTV planets are orange dashed).  The variable on the horizontal axis is the difference between the simulated or detected mass and the expectation value for that mass from the \citep{Wolfgang:2015} distribution ($M \sim 2.7 R^{1.3}$).  The thin dotted, blue curves correspond to the distribution when the RV detection threshold is modified to 1 and 0.5 m/s---the smaller this threshold, the more the distribution resembles both the TTV distribution and the true distribution.\label{cdfdiff}}
\end{figure}

\begin{figure}
\includegraphics[width=0.49\textwidth]{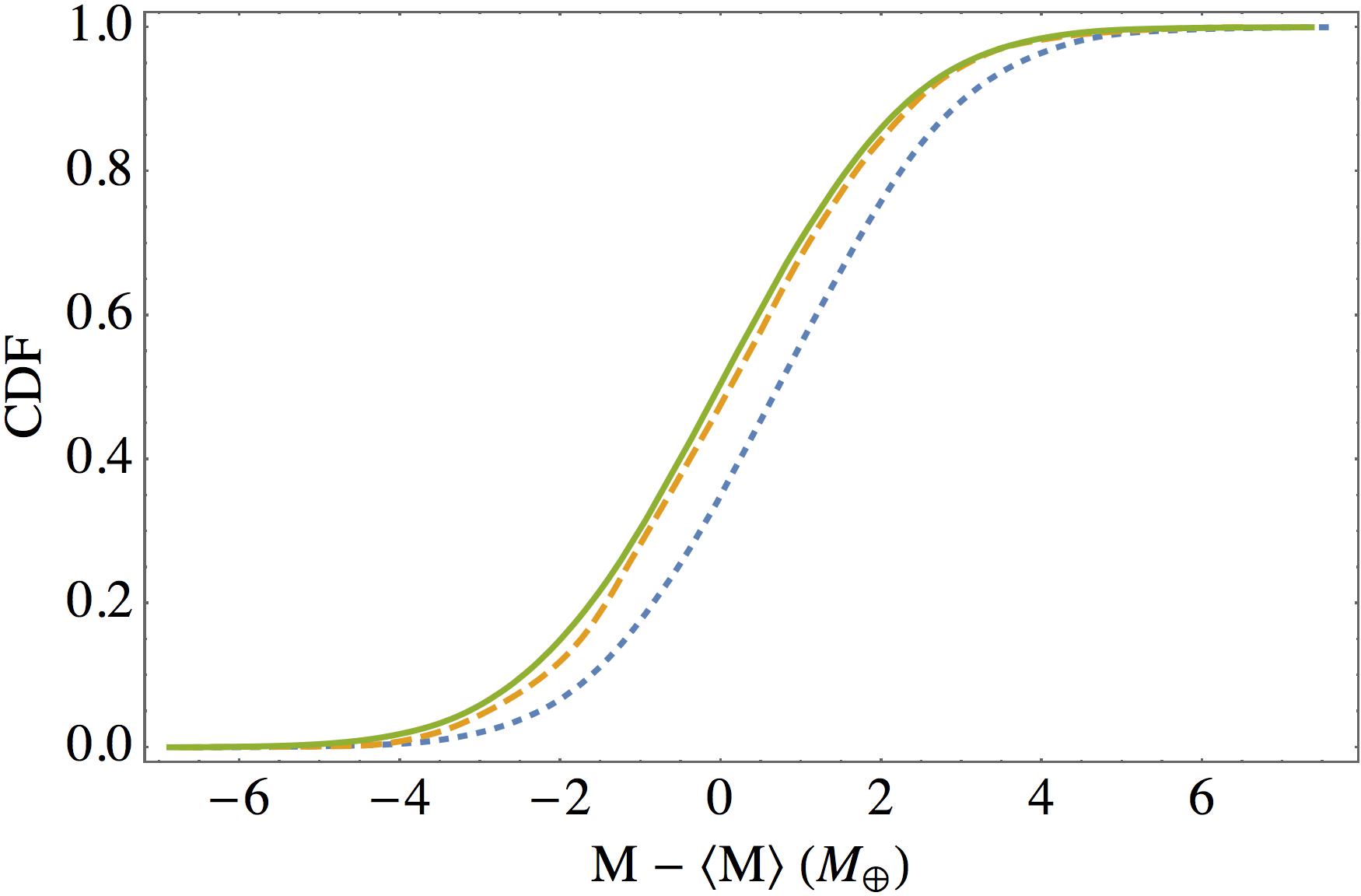}
\caption{Cumulative distribution of the simulated (green solid) and detected planets from our simulations (RV planets are blue dotten while TTV planets are orange dashed) for small planets with radii between 1.5 and 4 $R_\oplus$.  The variable on the horizontal axis is the difference between the simulated or detected mass and the expectation for that mass using the \citep{Wolfgang:2015} mass-radius distribution ($R \sim 2.7 R^{1.3}$).  We note that in this regime, the smallest mass planets are more difficult to measure with RV data than with TTV data and the TTV distribution more closely matches the underlying population.\label{cdfdiffsmall}}
\end{figure}

The differences in how the RV and TTV SNR vary with planet size, mass, and orbital period result in the two methods preferentially applying to different regions of the planet population parameter space.  This difference in the regions investigated by the two methods is especially true early on in a scientific program (as they currently tend to function) where high SNR targets are preferentially selected for additional scrutiny---and the differences in the sampled portions of the population become more exaggerated.  In order to best see the differences in the planet samples detected by these two methods, we look at the deviations of their masses from the expected value given by our chosen mass-radius distribution (i.e., $M - \langle M \rangle = M - 2.7 R^{1/3}$ with $M$ and $R$ in Earth units).  Also shown are the distributions with a 1 m/s and 0.5 m/s threshold for RV detections.  As the RV detection threshold drops, the distribution converges as expected toward both the TTV and true distributions.  Figure \ref{cdfdiff} shows the CDF of this quantity for the planet detections and the underlying population.  Figure \ref{cdfdiffsmall} shows the same distribution, but restricting the planet sizes to those between 1.5 and 4.0 $R_\oplus$.

\section{Discussion}\label{discussion}

Interestingly (surprisingly to the author), we see from both Figures \ref{cdfdiff} and \ref{cdfdiffsmall} that the bias from RV detections is larger than that from TTV detections.  This bias stems primarily from the fact that for a given radius, TTVs can detect a wider range of masses---probing to planets with smaller densities.  Thus, if forced to choose between methods (which we do not advocate), a complete TTV census would likely yield a better mass-radius relationship in the regime where planets transition from terrestrial to gaseous planets---the regime where most \kepler\ planets lie.

The fact that a TTV census would be less biased, and more sensitive to smaller planets, should be carefully considered when choosing an observing strategy for a transit survey.  If the primary goal of a transit survey is to discover a large number of planets, then short observations of many fields of view would be advantageous.  The number of planet discoveries would grow with the number of fields---though the discoveries would be limited to those with short orbital periods.  If, however, one wants to measure the masses of the discovered planets, to understand how planet properties change with distance from the host star, or to characterize the whole planetary system, then longer observations of a single field of view is preferable since the information in the TTV signal grows rapidly with a longer time series.

In particular, in the interval between the initial manifestation of the TTV signal until that signal completes an entire TTV cycle, the sensitivity to planet mass grows with $\sim t^{5/2}$ (the signal grows as $t^3$---the second term in the Taylor expansion of the sine function---while the noise grows as $t^{1/2}$).  If planet characterization via mass measurement is a primary science driver, then a continuous time series of a single field of view is nearly three times more valuable than two fields being observed for half as long ($2^{5/2}/2 \simeq 2.82$).  And, for the foreseeable future, TTVs are the only method capable of measuring the masses of the smallest planets out at moderate orbital distances (e.g., in or near the habitable zone of FGK dwarfs).

There are compelling reasons to pursue mass measurements using both RV and TTV data.  Among the most important is the possibility of populations of planets that have dynamical histories different from the typical system and which yield planets that are difficult to detect with one or the other method.  For example, any process that excites a planet's inclination makes TTV mass measurements much more challenging since it is unlikely that all of the planets that contribute to the TTV signal will transit the star.  Consequently, the analysis is exposed to significant degeneracies in the signal for different MMRs.  Therefore, a population of systems with large mutual inclinations requires RV measurements to fully characterize.

\begin{figure}
\includegraphics[width=0.49\textwidth]{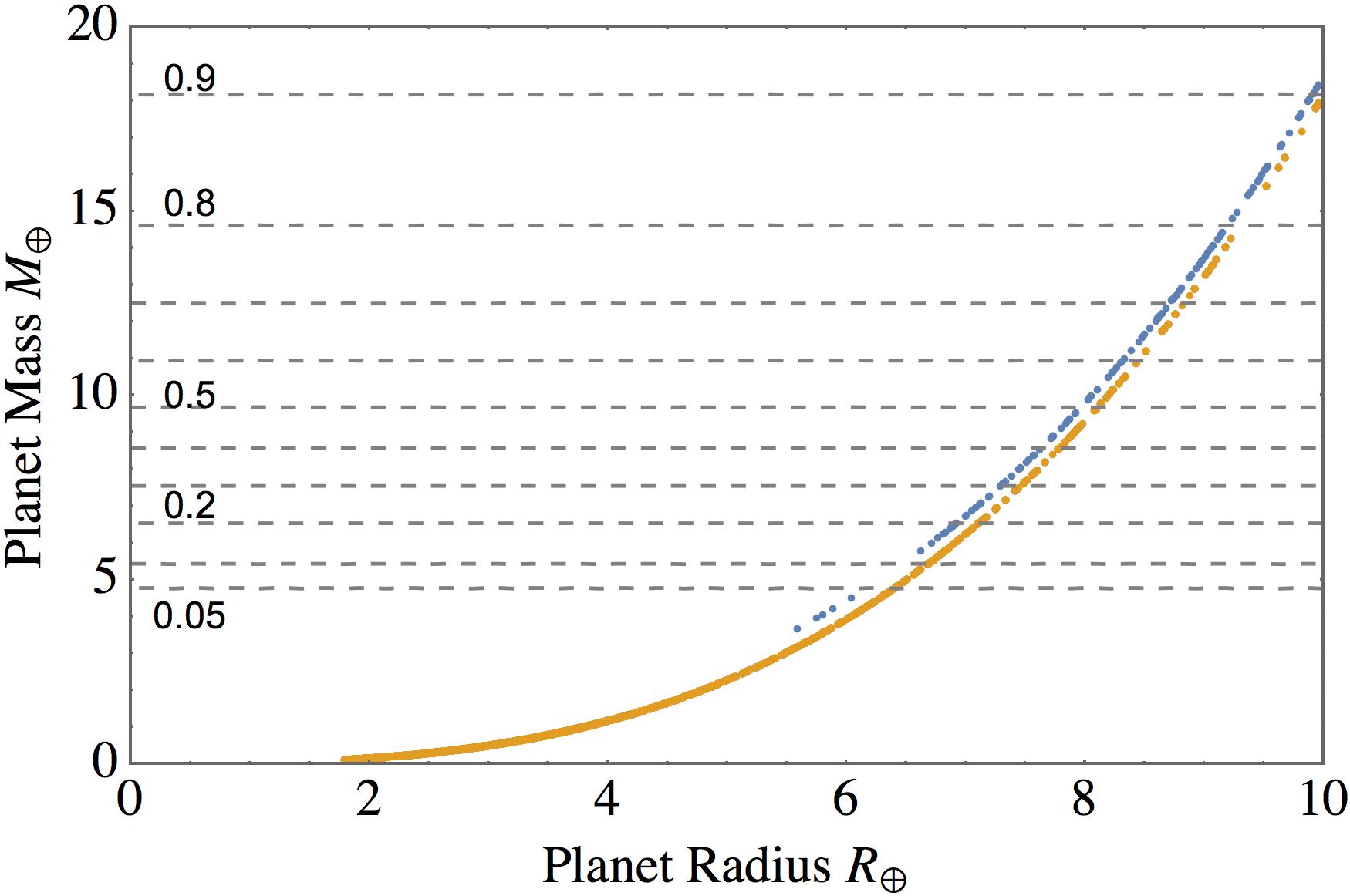}
\caption{Mass and radius of planets with a constant density of 0.1 g/cm$^3$ that are detected using RV (blue) and TTV (orange) data.  RV detections are vertically displaced by 0.5 $M_\oplus$ to make the graph more legible.  Here, RVs are only able to measure masses of the largest of these planets---providing only upper limits for smaller ones.  If nature only produces the small variety of these planets, then TTVs are the best means to measure their masses.\label{lodenseplot}}
\end{figure}

At the same time, a population of planets that have particularly low densities will be difficult to measure with RVs, yet may have a large TTV signal.  Figure \ref{lodenseplot} shows the detections for a population of planets with a uniform density of 0.1 g/cm$^3$.  Most of these planets lie well below the predictions of the \cite{Wolfgang:2015} mass-radius distribution, but we do observe planets with such low densities \citep[e.g.,][]{Masuda:2014,Jontof-Hutter:2014}.  In that figure we see that only the largest of these low-density planets are seen with RVs while TTVs can detect them down to masses much smaller than that of the Earth.

Ultimately, a number of effects may contribute to the difference between the planets whose densities are measured by TTVs and RVs.  Most of this paper was devoted to issues relating to the dependence of the SNR of the two methods on the orbital and physical properties of the planets.  Yet other physical effects may induce changes in otherwise identical planets.  We identify two here which merit further investigation.  One is any correlation between planet size and the flux it receives from the host star.  The other is a potential relationship between a planet's properties and the likelihood that it resides in or near MMR.

\begin{figure}
\includegraphics[width=0.49\textwidth]{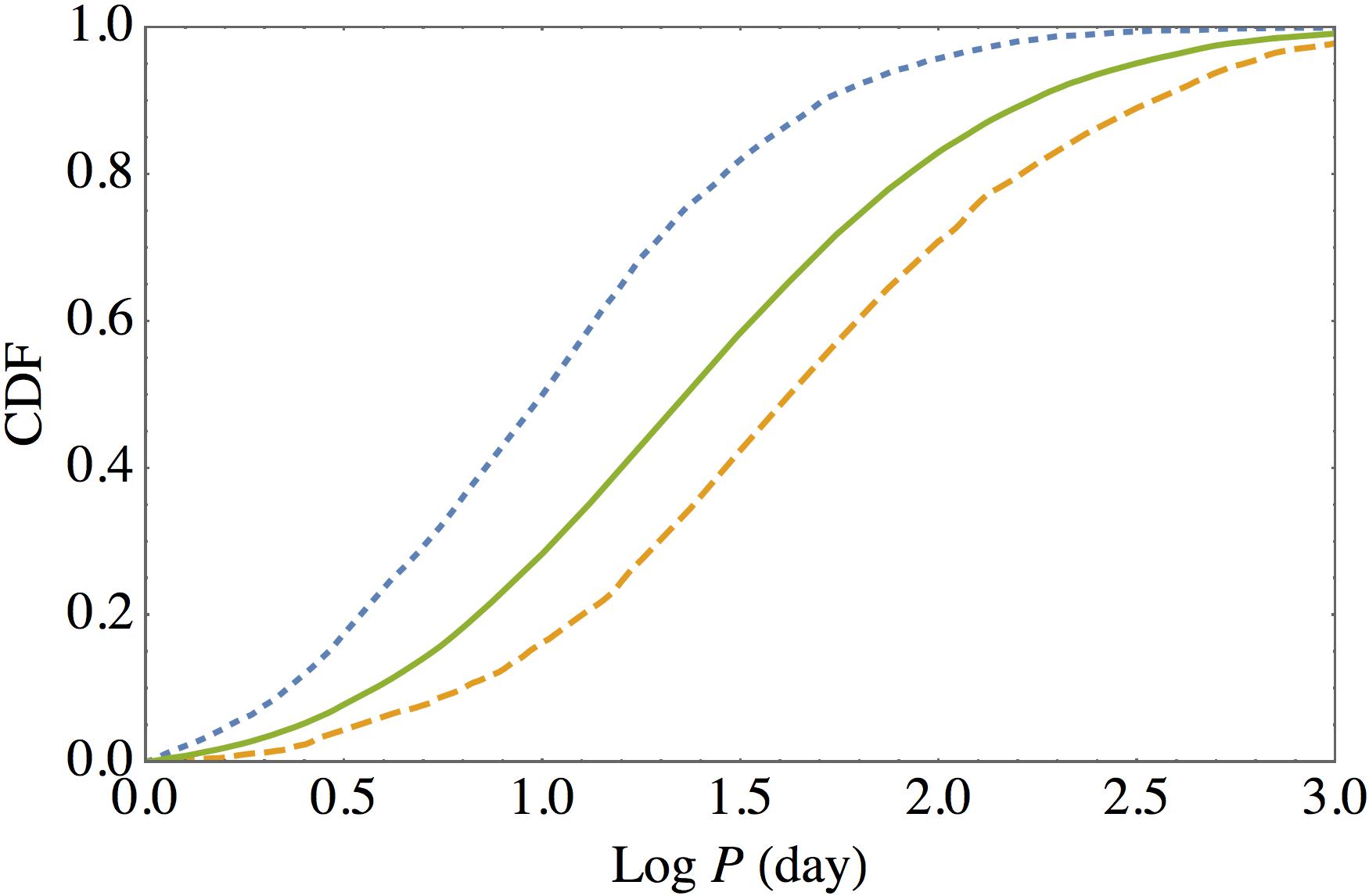}
\caption{Cumulative distribution of orbital periods of simulated (green solid) and detected planets from our simulations (RV planets are blue dotted while TTV planets are orange dashed).  This figure shows that RV measurements are more sensitive to planets on shorter orbital periods while TTV planets are more sensitive to longer orbital periods.  We do not consider here issues related to fewer transits at longer orbital periods, which reduces the effectiveness of TTV estimates.  One consequence of this difference in sensitivity is that any difference in planet structure due to proximity to the host star (e.g., systematically smaller planets due to evaporated atmospheres) would exacerbate the biases in the measured mass-radius relationships for each detection method.\label{perdetected}}
\end{figure}

Planets orbiting close to their star may have smaller atmospheres and consequently higher densities.  Since the RV signal is stronger at shorter orbital periods it would favor detecting planets with higher densities than either the progenitors of the observed planets or of planets with similar origins but with longer orbital periods.  Figure \ref{perdetected} shows the cumulative distribution of orbital periods for the initial and detected planets and indicates the systematic differences in orbital periods that are probed by the two methods.

\begin{figure}
\includegraphics[width=0.49\textwidth]{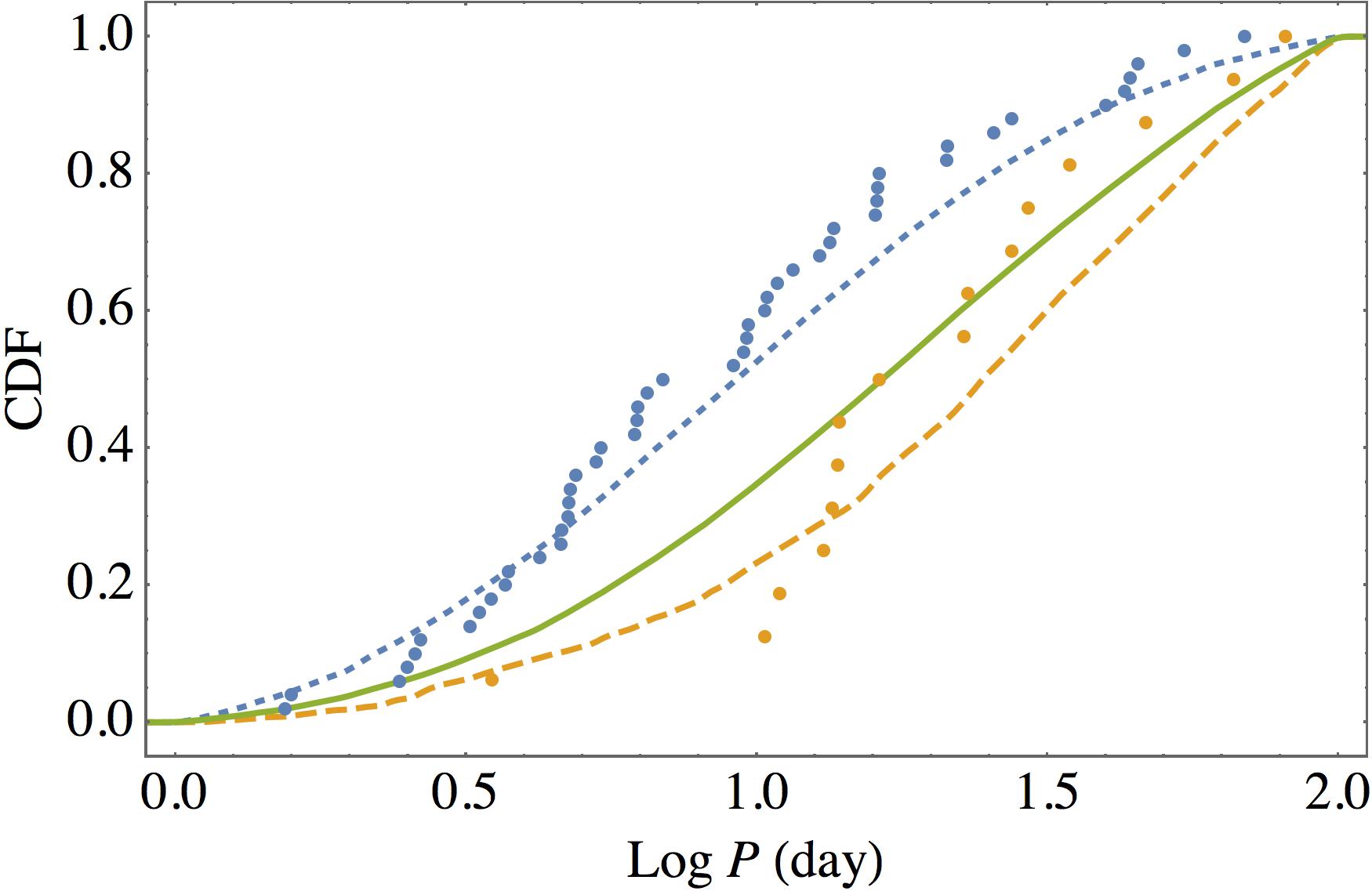}
\caption{Cumulative distribution of orbital periods of systems detected from our simulations and actual exoplanet observations with orbital periods from 1 to 100 days given in \citet{Wolfgang:2015}.  RV planets are blue (dotted) while TTV planets are orange (dashed).  The solid green line corresponds to the true distribution.  Even with a small sample, and with a heterogeneous target selection criteria, the differences in the distributions are clear and are consistent with the predictions of our model.  Anderson-Darling tests comparing the data with its corresponding model prediction yield $p$-values of 0.42 and 0.44 respectively for TTVs and RVs.\label{obsvssimperiod}}
\end{figure}

To illustrate this orbital period dependence, we select the planets listed in \citep{Wolfgang:2015} that were detected with RVs and TTVs having orbital periods between 1 and 100 days.  There are 50 planets detected with RVs and 16 with TTVs (this comprises all of the TTV systems and 50 of the 56 RV systems).  We note that only one of the 56 RV systems has an orbital period over 100 days while five (nearly 10\%) of the systems have orbital periods less than 1 day (55 Cnc at 0.74 days, CoRoT-7 at 0.85, and Keplers-10, 78, and 407 at 0.84, 0.35, and 0.67 days respectively).  By contrast, the shortest-period TTV planet is at 3.5 days.  Figure \ref{obsvssimperiod} shows the CDF of this sample of observed systems with the simulated data satisfying the same 1 to 100-day orbital period criterion.  An Anderson-Darling test comparing the simulated with the observed systems yields $p$-values of 0.42 and 0.44 for the TTV and RV samples respectively.  Thus, even in the absence of specific effort to produce a complete survey (especially of TTV systems) the dependence on orbital period is manifest and agrees (i.e., doesn't disagree) with the model predictions.

Given this bias in sensitivity to planetary orbital periods, if there is a strong correlation between planet density and orbital period---especially near the host star---then we would expect an exaggerated bias from the RV sample.  At the same time, planet pairs near first-order MMR may have a history of strong or exotic dynamical interactions---including possible couplings to the atmosphere (spin-orbit) that could significantly affect the apparent size of the planets.  Such a scenario could yield an overabundance of planet mass measurements through TTVs if the planets near MMR are systematically larger since the larger planet sizes reduce the timing uncertainty.  The true nature of the atmospheric differences between planets near MMR or near the host star is left for other work as a number of subtle effects must be considered.  For example, if RV planets migrated closer to the host star than TTV systems, it may imply that they formed from a more massive disk---which would generally produce larger planets thereby negating some of the effects of evaporation in the early lives of the systems.  Both enumerating and accounting for the wide variety of possible effects is nontrivial and will require detailed investigations.

We note that our study does not consider issues relating to the detection of the planet transits in the first place.  For planets that are sufficiently small as to be missed by the transit survey, neither the RV nor the TTV method can yield a datum on the mass-radius distribution.  The planet could, in principle, be detected by either method through its effect on the star or other planets in the system.  However, the lack of a transit signature means that the RV search would be blind and the TTV search largely degenerate.  Thus, for the smallest and least massive planets, only new technologies in instrumentation can enable probes of that region of parameter space.

Regardless of the details, in order for us to understand the formation and physical histories of exoplanets, we need to employ, in a systematic way, all of the tools available to us.  Based on our simulations, we find the importance of TTV mass measurements to be clear---especially as it pertains to planets with smaller sizes $\lesssim 2 R_\oplus$ and at orbital periods beyond a few days.  Nevertheless, a comprehensive study using both methods is the only means to empirically, and reliably, measure the properties of exoplanets and their abundances spanning the observed distributions of orbital period and system architecture.

\section*{Acknowledgements}
J.H.S. acknowledges support from NASA under grant NNH12ZDA001N-KPS 
issued through the Kepler Participating Scientist Program and grant NNH13ZDA001N-OSS issued through the Origins of Solar Systems program.  He thanks the multibody working group for the \kepler\ mission for stimulating conversations that enriched this work, especially the constructive comments from Eric Ford and Jack Lissauer.  He is especially grateful to those who shared their ongoing work with him (Kat Deck, Eric Agol, Tomer Holczer, and Tsevi Mazeh).  This research has made use of the NASA Exoplanet Archive, which is operated by the California Institute of Technology, under contract with the National Aeronautics and Space Administration under the Exoplanet Exploration Program.

\bibliographystyle{plainnat}
\bibliography{multis}

\bsp

\label{lastpage}

\end{document}